\documentclass[prl,10pt,twocolumn]{revtex4}

\usepackage{amsfonts}
\usepackage{amssymb}
\usepackage{dsfont}
\usepackage{amsmath}
\usepackage{graphicx}
\usepackage{multirow}

\newcommand{\be}{\begin{equation}}
\newcommand{\ee}{\end{equation}}
\newcommand{\ba}{\begin{eqnarray}}
\newcommand{\ea}{\end{eqnarray}}
\newcommand{\n}{\nonumber}
\newcommand{\lan}{\langle}
\newcommand{\ran}{\rangle}
\newcommand{\B}{\beta}
\newcommand{\A}{\alpha}

\begin{document}
\title{Near-unit fidelity entanglement distribution using Gaussian communication}

\author{Ludmi{\l}a Praxmeyer and Peter van Loock}

\affiliation{
Optical Quantum Information Theory Group, Max Planck Institute for
the Science of Light, Institute of Theoretical Physics I,
Universit\"{a}t Erlangen-N\"{u}rnberg, Staudtstr.7/B2, 91058
Erlangen, Germany}

\begin{abstract}
We show how to distribute with percentage success probabilities
almost perfectly entangled qubit memory pairs over repeater
channel segments of the order of the optical attenuation distance.
In addition to some weak, dispersive light-matter interactions,
only Gaussian state transmissions and measurements are needed for
this scheme, which even beats the coherent-state-benchmark for
entanglement distribution based on error-free non-Gaussian
measurements. This is achieved through two innovations: first,
optical squeezed states are utilized instead of coherent states.
Secondly, the amplitudes of the bright signal pulses are
reamplified at each repeater station. This latter variation is a
strategy reminiscent of classical repeaters and would be
impossible in single-photon-based schemes.
\end{abstract}

\maketitle

The maximum distance for experimental quantum communication is currently about 250 km
\cite{stucki09,QKDRMP}. Although extensions to slightly larger distances are possible
based on present experimental approaches \cite{scheidl09},
truly long-distance quantum communication, similar to classical communication
networks on inter-continental scale, would require turning the theoretical in-principle
solution of a quantum repeater \cite{nn,Innsbruck} into a real implementation \cite{QRRMP}.
This, however, would be possible only provided that highly sophisticated
subprotocols such as efficient entanglement distillation \cite{standard1,standard2}
and at the same time sufficient quantum memories \cite{Memory} are within experimental reach;
only with these extra ingredients can we circumvent the otherwise exponential decay
of either communication rates or fidelities in the presence of channel losses.

There are several proposals for implementing a quantum repeater
\cite{nn,Innsbruck,Duan01,Harvard,PvL}, utilizing different physical systems, and
varying in their consumption of spatial versus temporal resources.
In all these schemes, some kind of heralding mechanism is needed
in order to conditionally distribute entangled pairs between
neighboring repeater stations. Among other classifications, for
our purpose, it is useful to divide these schemes into two
categories: one, where single photons are used to distribute
entanglement, and another one, where bright optical coherent
states are exploited (``hybrid quantum repeater" \cite{PvL}, HQR).
In the former class of repeaters, as vacuum contributions and
photon losses would mainly affect the distribution efficiencies
and not the quality of the created pairs, the heralding
probabilities are typically fairly low, but initial fidelities are
naturally quite high. Conversely, the quality of the
bright-light-based pair distribution is very sensitive to losses;
hence fidelities are modest, but postselection efficiencies are
reasonably high.

For realizing a full quantum repeater, however, it is a priori not obvious which approach
is preferable (especially, when imperfect quantum memories are considered):
that leading to high-fidelity initial entanglement at low rates or that based upon
higher initial distribution efficiencies at the expense of lower initial fidelities.
Nonetheless, in general, the globally optimal quantum repeater protocol
(achieving a certain target fidelity for long-distance pairs at an optimal rate) would always
combine optimal subprotocols for entanglement distribution, distillation, and connection
\cite{JiangDynProgr}. Hence distribution of entangled pairs between neighboring stations should
occur at an optimal rate for a whole range of useful short-distance fidelities. This {\it tunability}
of optimal efficiency versus fidelity, and, in particular, {\it near-unit fidelity} pair distribution
is impossible to obtain in the HQR scheme based on coherent states and homodyne
detection \cite{PvL}.

There were several proposals for modifying the original HQR scheme,
mainly differing in the type of measurements
used. These variations then do allow for tunability and near-unit fidelity entanglement
distribution; but the required POVMs involve experimentally demanding non-Gaussian
detection schemes such as cat-state projections \cite{Munro_Nemoto}, photon-number
resolving detectors or, at least, detectors discriminating
between vacuum and non-vacuum states \cite{PvL_3,jap2}.
A benchmark on the fidelity versus success probability plane can be derived,
based upon the non-Gaussian POVM achieving optimal, error-free unambiguous state discrimination
(USD) of coherent states \cite{PvL_3}. This benchmark covers the whole range of
useful fidelities and it can be approached or even attained through non-Gaussian
photon detectors \cite{PvL_3,jap2}.

\begin{figure}[t]   
$\!\!\!$\includegraphics[scale=0.39]{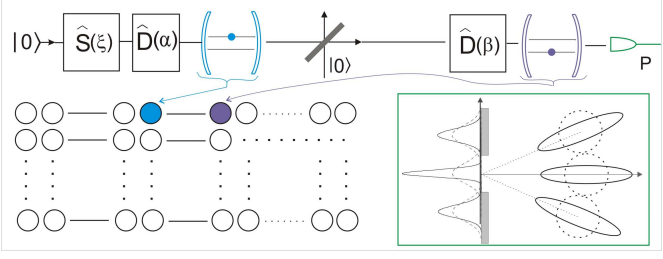}
\caption{Entanglement distribution
using squeezed-state ($|\A,\xi \ran$) communication with
reamplification ($\hat{D}(\B)$). Reamplified phase-rotated
squeezed states can be better discriminated through homodyne
detection than unamplified coherent states.}\label{squeeze}
\end{figure}

In the present work, we address the question whether it is possible to switch back
from the rather demanding and less practical
non-Gaussian schemes to a scheme fully based on {\it Gaussian resources and operations}
without loss of performance.
We answer this question to the affirmative and, in particular,
we show that even the coherent-state USD benchmark
can be beaten in a Gaussian protocol that allows for just the right amount
of measurement-induced overlap errors.
For this we introduce two innovations involving Gaussian
resources: the use of optical squeezed states
instead of coherent states; and the reamplification of the signal amplitude at
each repeater station. Squeezing 
improves the distinguishability
of the final states along certain directions in phase space, see Fig.~\ref{squeeze}.
Reamplification is a strategy
reminiscent of classical repeaters -- a modification that would be
impossible in single-photon-based schemes.

Here we optimize
Gaussian communication for the HQR scheme with the only restriction that
the initial probe beam is in a pure state, and under the natural
assumption that the initial squeezing direction and the final
quadrature projection axis coincide.
We will combine ingredients already exploited in
Ref.~\cite{PvL}, i.e., dispersive atom-light interactions, a beam
splitter loss model, and homodyne detection with squeezing and
reamplification.
{\it Ideal qubit-qumode interactions--} For the initial entanglement distribution
in an HQR, two neighboring stations are each
equipped with a cavity
containing a two-level system (qubit)${}^{\footnotemark}$ and connected
by a channel that can carry a quantized optical mode (qumode).
\footnotetext{as for the discrete
spin variables we may refer to ``atoms", although these could as well
be quantum dots, donor impurities in semiconductors, etc.}
The qumode is initially in a displaced squeezed vacuum state,
$|\alpha,\xi\ran= \hat{D}(\A) \hat{S}(\xi) |0\ran$, with $\xi=r
e^{i\pi}$, real parameters $\A$ and $r$, and the displacement and
squeezing operators $\hat{D}(\A)$ and $\hat{S}(\xi)$, respectively
\cite{sq}.
The initial atomic states are each
$(|0\ran+|1\ran)/\sqrt{2}$. Now a dispersive off-resonant
interaction,
$\hat{U}_{int}=\exp(i\theta
\hat{n}\sigma_z/2)$, on the first qubit, where $\hat{n}$ is the photon
number and {\mbox{$\sigma_z=|0\ran\lan 0|-|1\ran\lan
1|$}} the Pauli $Z$ operator, leads to a conditional phase-rotation of the qumode,
$\hat{U}_{int}
\left[\frac{|0\ran+|1\ran}{\sqrt{2}}\otimes |\A,\xi\ran
\right]= \frac{|0\ran\otimes |\A_0, \xi_0\ran+|1\ran\otimes |\A_1,
\xi_1\ran }{\sqrt{2}}$
with {\mbox{$\A_k= e^{\frac{i\theta(-1)^k}{2}} \A$}},
{\mbox{$\xi_k=e^{(-1)^k i\theta }\xi $}}, where {\mbox{$k=0,1$}}.
After this first interaction, the qumode travels to the other cavity
and interacts with the second qubit in a similar way. For a
loss-free channel, the final qubit--qubit--qumode state is given by
$\left[ |0\ran |0\ran |e^{i\theta}
\A,e^{2 i
\theta}\xi\ran +|1\ran |1\ran
|e^{-i\theta} \A,e^{-2 i \theta}\xi\ran\right.
+\left.\big(|0\ran |1\ran+|1\ran |0\ran\big)| \A,\xi\ran
\right]/2$.
By measuring the qumode in an appropriate way
one can distinguish its initial state $|
\A,\xi\ran$ from the phase-rotated states and
conditionally create an entangled state
between the two cavities \cite{PvL}.


{\it Lossy channels--} In the realistic scenario, two neighboring
repeater stations are separated at least by a distance of $10-20$ km,
linked by a lossy channel of this length. Thus, the qumode will be subject
to attenuation and thermalization, especially when its initial state differs from a pure
coherent state.
In order to describe the resulting mixed-state density matrices,
we define the operator
$\hat{L}_{jk}$ [see Eq.~(\ref{sq_N}) in Appendix ``Methods"];
it characterizes our system after the interaction in
the first cavity and the transmission through the lossy channel
(derivation of $\hat{L}_{jk}$ and
more details about the noise model can also be found in the Appendix).
\noindent

While decoherence or thermalization are unavoidable in the lossy channel,
the effect of attenuation may be corrected by an additional displacement operation
$\hat{D}(\B)$. Thus, before the interaction in the second
cavity,  
we displace the light field by a suitably chosen, real $\B$.
The total state of the system (qumode and two qubits) after the
second interaction is given by
\ba
\hat{\rho}\!=\!\!\!\displaystyle\sum_{l,m,j,k=0}^{1}
\!\!\!|j,l\ran\lan k,m| e^{\frac{i\theta \hat{n}}{2}(-1)^{l} }
\hat{D}(\B) \hat{L}_{jk} \hat{D}^\dagger(\B)   e^{\frac{i\theta \hat{n}
}{2} (-1)^m }\!\!,\label{eq1}
\ea
where the $l,m$ indices label the atomic states in the
second cavity, while the indices $j,k$ refer to the
first cavity. Now measuring the qumode subsystem of $\hat{\rho}$
leads to a conditional {\mbox{4 by 4}} two-qubit density matrix.
In the case of homodyne detection of the $p$
quadrature,
i.e., $\displaystyle {{\int_{-p_c}^{p_c}}}\!
\lan p|\hat{\rho} |p\ran\, dp$,
we effectively select from $\hat{\rho}$ those terms corresponding
to a mixture of $|\Psi^\pm\ran=(|01\ran\pm|10\ran)/\sqrt{2}$ Bell
states; the resulting phase-flip errors ($\pm$) stem from photon
losses, minimal for small amplitudes $\alpha$; the finite overlaps
of the Gaussian peaks in the homodyne-based approach lead to
additional bit-flip errors, minimized for large amplitudes
$\alpha$
\cite{PvL_3}. However, in our generalized scheme, we have as
additional parameters the squeezing $r$ and the reamplification
amplitude $\beta$, which have a significant impact on the above
trade-off between channel decoherence and homodyne-based
Gaussian-state distinguishability.

The final fidelity compared to
the ideal Bell state $|\Psi^+\ran$
now becomes
$ F= \displaystyle\int_{-p_c}^{p_c}\! \lan
\Psi^+|\lan p|\hat{\rho} |p\ran |\Psi^+\ran\, dp/P_s\,.$ 
\noindent
The normalization factor, $P_s= \mathrm{Tr}\left[
\displaystyle\int_{-p_c}^{p_c}\!\lan p| \hat{\rho}|p\ran \,
dp\right]$, after tracing over the conditional qubit states, determines
the probability of success, i.e., how frequently we actually
obtain a measurement result within the postselection window $2 p_c$.
Exact expressions for $F$ and $P_s$ are given in the Appendix
[Eqs.~(\ref{Fid}), (\ref{Ps})].

\begin{figure}[h]  
\includegraphics[scale=.54]{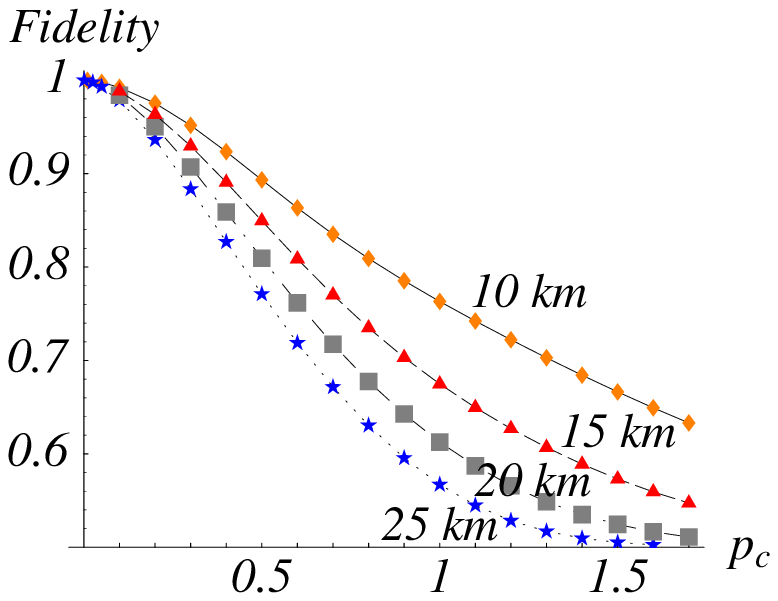}\includegraphics[scale=.54]{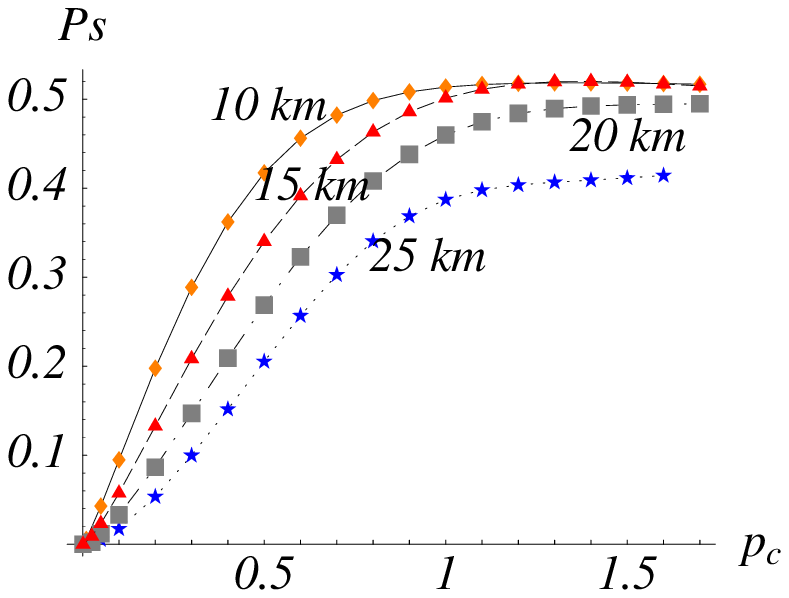}
\caption{Fidelity $F_{Abs}$ and corresponding probability of
success $P_s$ as a function of the selection window width $p_c$
for transmission distances 10, 15, 20, and 25 km,  rotation angle
$\theta=0.01$ and loss 0.17 dB/km. Free parameters used for
optimization were: initial $\A$, squeezing parameter $r$, and
displacement $\B$. }\label{fig1_multi}
\end{figure}

{\it Results--} Although the fidelity $F$ [Eq.~(\ref{fid})] is a highly
oscillating function, we shall focus on its upper envelope
$F_{Abs}$, calculated from $F$ by taking the absolute value instead of
the real part of the last term in Eq.~(\ref{fid}),
as we may always ``undo'' the corresponding local phase (see
Refs.~\cite{PvL,Ladd06} for details).
From now on we assume fixed phase shift $\theta=0.01$ and transmission $T$,
with losses corresponding to 0.17 dB/km.
The fidelity then becomes a function of the squeezing parameter $r$, the initial
amplitude $\A$, the displacement $\B$, and the selection window $2 p_c$. Varying
$\A$, $\B$, $r$ for every $p_c$, the maximum of $F_{Abs}$ can be
found. The plots in Fig.~\ref{fig1_multi} show
the maximal $F_{Abs}$ with corresponding
$P_s$ for different distances between repeater stations.
We see that now near-unit fidelities can
be achieved owing to squeezing {\it and}
reamplification: for the selection window $p_c \rightarrow 0$,
the maximal fidelities approach unity, at the expense of
success probabilities tending to zero. This regime was previously
accessible only through non-Gaussian measurements such as USD or in
conceptually different single-photon-based schemes.

\begin{figure}[h]   
\includegraphics[scale=.7]{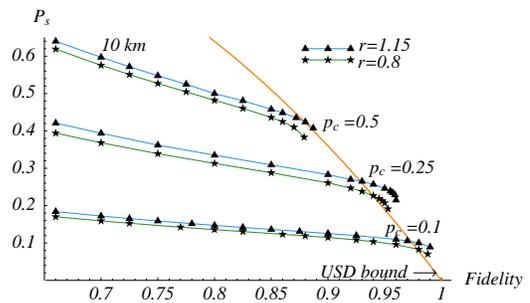}
\caption{Maximal probability of success $P_s$ for given
$F_{Abs}$ and different $p_c$ and $r$  (rotation angle is set to
$\theta=0.01$, loss to 0.17 dB/km, and distance $L_0$ is 10 km).
Maximal fidelity $F_{Abs}$ obtained choosing optimal initial $\A$
and displacement $\B$ for  $r=0.8$ and $r=1.15$ and $p_c=0.1$,
$p_c=0.25$, $p_c=0.5$, respectively, is shown. Orange curve is the
USD bound
\cite{PvL_3} valid for coherent states. }\label{fig3}
\end{figure}
\begin{figure}[h]
$\!\!\!\!\!\!$\includegraphics[scale=.85]{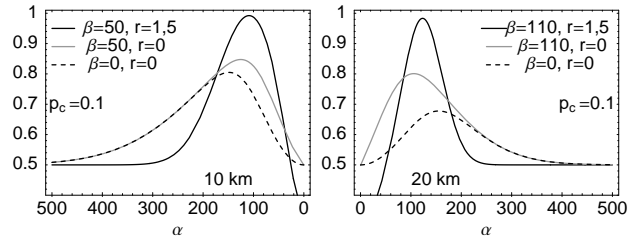}
\vspace*{-0.5cm}
\caption{Fidelity  $F_{Abs}$ as a function of initial $\A$.
 Dashed lines correspond to a coherent state as initial qumode state and no
amplification, solid grey lines to a coherent state amplified,
solid black lines to a squeezed state ($r=1.5$) amplified after
transmission. Distance between cavities is $L_0=10$ or $20$ km,
and amplification is just a displacement by $\hat{D}(\B)$.
}\label{figA.0}
\end{figure}

Alternatively, we may obtain maximal
$P_s$ for fixed F, $p_c$, and $r$, see
Fig.~\ref{fig3} with $0.66\leq F_{Abs}\leq 1$, $L_0=10 $ km,
$p_c\in\{0.1,0.25,0.5\}$, and  $r\in\{0.8,1.15 \}$.
For comparison we included the coherent-state USD bound
\cite{PvL_3} (in orange), previously obtainable only through non-Gaussian POVMs
\cite{jap2}. We observe that for sufficiently small selection windows,
our scheme combining squeezed light, reamplification, and homodyne
detection performs better then those based on single-photon
detectors. Similar but slightly smaller improvements
over the USD bound can be obtained for a distance of the order of the
attenuation distance, $L_0=20$ km.
A comparison of the standard HQR scheme \cite{PvL} and ours with squeezing
and reamplification 
is given in Fig.~\ref{figA.0} and Table~\ref{tab1}.
The differences are significant. For $L_0=10$ km, both
fidelities and probabilities of success are much higher in
our scheme; for $L_0=20$ km, fidelities are highly increased, at
the expense of smaller success probabilities.

\begin{table}[ht]
\caption{Fidelities $F_{Abs}^{max}$ and probabilities of success
$P_s$  corresponding to parameters from Fig.~\ref{figA.0} }
\label{tab1}
\begin{tabular}{|c|l||c|c||c|c|c|c|}
\hline \multicolumn{2}{|c||}{initial state
}&\multicolumn{2}{|c||}{squeezed,}& \multicolumn{2}{|c|}{coherent,
}&\multicolumn{2}{|c|}{coherent,
no } \\
\multicolumn{2}{|c||}{of light:
}&\multicolumn{2}{|c||}{amplified}& \multicolumn{2}{|c|}{amplified
}&\multicolumn{2}{|c|}{amplification} \\
\hline distance & $\;p_c\;$  & $F_{Abs}^{max}$ & $P_s$ [\%] & $F_{Abs}^{max}$ &
$P_s$ [\%]& $F_{Abs}^{max}$ & $P_s$ [\%]\\
\hline
\multirow{3}{*}{10 km}& $\;0.1\;$ & 0.99 & 9 & 0.85 & 7 & 0.80& 8  \\
&$\;0.25\;$ & 0.96 & 23 & 0.83 & 18 & 0.80& 20\\
&$\;0.5\;$ & 0.89 & 40  & 0.80 & 33 & 0.77& 36\\
  \hline
\multirow{3}{*}{20 km}& $\;0.1\;$ & 0.98 & 4.5 & 0.79 & 5 & 0.68& 9 \\
&$\;0.25\;$ & 0.93 & 12 & 0.77 & 13 & 0.67 & 21\\
&$\;0.5\;$ & 0.81 & 26 & 0.71 & 26 & 0.63 & 39\\
  \hline
\end{tabular}
\end{table}


{\it Discussion--} The results presented here are restricted to an
elementary segment of a full HQR. Obviously, the present scheme
gives a lot of freedom regarding optimal fidelity and probability
of success, as a starting point for the subsequent procedures of
entanglement swapping and purification. Even though in our
generalized scheme, initial fidelities are high, we should stress
that the resulting two-qubit entangled-state density matrices have
non-zero elements for all four Bell states, as opposed to, for
instance, the non-Gaussian USD-based scheme \cite{PvL_3}.
%
The rank-2 mixtures there \cite{PvL_3} are typically easier to
purify than the full rank-4 mixtures obtained from both
photon-loss-induced phase-flip and measurement-induced bit-flip
errors as in our scheme. We leave a full analysis, incorporating
our scheme into a complete HQR including rank-4 purifications and
swappings for future research. The reason why in our scheme we can
suppress the loss-induced errors to a great extent is because we
may keep the initial amplitudes $\A$ relatively small, but still
have only small amounts of measurement-induced errors owing to
squeezing and reamplification.

We note that different from existing proposals for
distributing discrete entanglement through dynamical entanglement
transfer from two-mode squeezed
\cite{KrausCirac} or general two-mode states
\cite{Paternostro} to discrete systems,
our scheme makes explicit use of weak (dispersive, off-resonant)
light-matter interactions and employs local measurements including
postselection; photon losses are primarily assumed to occur in the
channel as a limiting factor to the communication distance,
instead of distance-independent dissipation during the local
interactions
\cite{KrausCirac,Paternostro,Chin}.




{\it Realistic qubit-qumode interactions--} Finally, we
address the question whether the idealized, controlled phase-rotation
in our scheme
can be indeed approximately realized;
especially, when the qumode starts in a nonclassical, squeezed state.
First of all, the
effective Jaynes-Cummings-based interaction for the limiting case
of large detuning in the off-resonant, dispersive regime holds for
any input state of the qumode. However, in a cavity-QED setting,
the internal cavity mode and the external fields are no longer
identical; in particular, atomic spontaneous emissions (unwanted in-out
couplings) and a finite desired cavity in-out coupling
have to be taken into account.
The master equation derived in Ref.~\cite{Ladd06} under the Born
approximation holds for any qumode state; in the relevant regime
of $\alpha$ values, semi-classical calculations are
sufficient, however, we have to assume that a
squeezed state coupled into the cavity at least remains a Gaussian
state at all times. As a result, non-Gaussian effects become
negligible, similar to the case of coherent-state inputs of
Ref.~\cite{Ladd06}. The crucial parameter is then a sufficiently
large cooperativity (``good coupling / dissipation")
at weak or intermediate coupling.

Squeezing may even turn out to
be beneficial for the fidelity of the dispersive
interaction~\cite{Chin2}. In our model, coupling inefficiencies
may be absorbed into the transmission parameter $T$, corresponding
to reduced distances. Alternatively, the optical
squeezing operation may be postponed
until the very end, performed online~\cite{Akira} on phase-rotated
coherent states.
Besides CQED,
approaches less sensitive to local dissipations may involve
free-space light-matter couplings~\cite{Savage,Sondermann}.


{\it Acknowledgements--} P.v.L. wants to thank Bill Munro and Kae Nemoto for
their support when this research line was initiated. The authors
acknowledge support from the Emmy Noether programme of the DFG in Germany.

\section{Methods}
{\small
\noindent
{\it Beamsplitter noise model}-- we assume that the incident light
mode $a$ interacts on a beamsplitter with an additional mode $b$
(initially in a vacuum state). After this interaction the trace over mode $b$
is taken, assuming no control over the loss mode.
A beamsplitter transforms two incident modes according to the
following unitary operation: $ 
\hat{U}^\dagger_{\mathrm B} \hat{a} \hat{U}_{\mathrm
B}=\hat{a}\sqrt{T} +i \,\hat{b} \sqrt{R}$, $\hat{U}_{\mathrm
B}^\dagger
\hat{b}\, \hat{U}_{\mathrm B}=\hat{b}\sqrt{T} +i
\,\hat{a}\sqrt{R}$, where the standard relation between the
reflection and transmission coefficients, i.e., $T+R=1$, holds.
Thus, the interaction of a displaced squeezed vacuum state with a vacuum state
on a beamsplitter leads to the following
state:
\ba
 &|\psi_{ab}(\A,\xi)\ran:=\hat{U}_{\mathrm B}\left[ \big(\hat{D}(\A) \hat{S}(\xi)
 \otimes \mathds{1}\big) |0_a;0_b\ran\right]=&\n\\
&=\frac{1}{\sqrt{\mu}} \exp\left(\A \big(\sqrt{T}\hat{a}^\dagger
+i
\sqrt{R}\hat{b}^\dagger \big)-
\A^*\big(\sqrt{T}\hat{a} -i\sqrt{R}\hat{b} \big) \right)&\n\\
& \exp\left[-\frac{\nu}{2\mu}\left(T\hat{a}^{\dagger 2}-R
\hat{b}^{\dagger 2}+2i\sqrt{RT} \hat{a}^\dagger \hat{b}^\dagger
\right)\right] |0_a;0_b\ran,&
\n
\ea
where $\mu=\cosh|\xi|$, $\nu=\xi/|\xi|\sinh|\xi|$.

Tracing over mode $b$, we find that the ``light'' part of an
arbitrary element of the density matrix
\ba
\sum_{j,k=0}^1 |j\ran\lan k|\otimes|\A_j, {\xi_j}\ran\lan
\A_k, {\xi_k}|\n
\ea
that corresponds to
\ba
&\hat{U}_{int}
\left[\frac{|0\ran+|1\ran}{\sqrt{2}}\otimes |\A,\xi\ran
\right]= \frac{|0\ran\otimes |\A_0, \xi_0\ran+|1\ran\otimes |\A_1,
\xi_1\ran }{\sqrt{2}}\,,\n
\ea
is transformed into:

\begin{widetext}
\ba & L_{jk}:=\mathrm{Tr}_b\big[|\psi_{ab}(\A_j,\xi_j)\ran
\lan\psi_{ab}(\A_k,\xi_k)| \big]=
e^{-\frac{|\A_j|^2+|\A_k|^2}{2} } 
 e^{\A_j\A_k^*
R}{\displaystyle{\iint}}
\frac{\mathrm d^2 \mathrm H}{\mu}
\frac{\mathrm d^2 \mathrm J}{\pi^2} e^{-|\mathrm H|^2-|\mathrm
J|^2+\frac{\nu_j R}{2 \mu} \mathrm J^{* 2}+\frac{\nu_k^* R}{2\mu}
\mathrm H^2+\mathrm H^*\mathrm J}\times\n\\
&e^{i \sqrt{R}(\A_j^*-\A_k^*)J}
 e^{ \A_j \sqrt{T}\hat{a}^\dagger} e^{-\A_j^* \sqrt{T} \hat{a}}
e^{-\frac{\nu_j T}{2 \mu}  \hat{a}^{\dagger 2} -i\frac{\nu_j
\sqrt{RT}}{ \mu}  \hat{a}^\dagger
 \mathrm J^*} 
  |0\ran\lan 0|
e^{i\frac{\nu_k^* \sqrt{TR}}{\mu}\hat{a}\mathrm  H-\frac{\nu_k^*
T}{2\mu} \hat{a}^2} e^{-\A_k \sqrt{T}\hat{a}^\dagger}e^{\A_k^*
\sqrt{T}\hat{a}}e^{i \sqrt{R}(\A_j-\A_k)
\mathrm H^*}
\label{sq_N}
\ea
where ${\mathrm d}^2 \mathrm  H=  {\mathrm d} (\mathrm{Re}
\mathrm H) \mathrm{d} (\mathrm{Im} \mathrm H) $, ${\mathrm d}^2
\mathrm J= {\mathrm d}(\mathrm {Re} \mathrm J)  \mathrm{d}
(\mathrm{Im}
\mathrm J$).

\noindent
{\it Fidelity, Probability of success}-- After the transmission
through a lossy channel the light is reamplified by applying a
displacement operator $\hat D(\B)$ to the qumode. Then the qumode
interacts with the atom in the second cavity, and is finally
detected by a homodyne measurement of the {\mbox{$p$-quadrature.}}
In our notation, we have $\hat{x}=(\hat{a}+\hat{a}^\dagger)/2 $,
$\hat{p}=(\hat{a}-\hat{a}^\dagger)/(2i) $; corresponding to a
commutator $[\hat{x},\hat{p}]=i/2$. The total state of the system
before measurement is described by $\hat{\rho}$ from
Eq.~(\ref{eq1}).
 The fidelity of the
(renormalized) conditional state measured within the postselection
window {\mbox{[-$p_c$, $p_c$]}}, compared to the ideal Bell state
$|\Psi^+\ran$, reads as follows:
\ba
\label{Fid}
&F=\frac{1}{P_s} \displaystyle\int_{-p_c}^{p_c}\! \lan
\Psi^+|\lan p|\hat{\rho} |p\ran |\Psi^+\ran\, dp=
\frac{1}{P_s}
\left\{ \mathrm{Erf}\left[\frac{\sqrt{2} (p_c-\B \sin\frac{\theta}{2})}{\sqrt{1+2T\nu(\nu+\mu)}}\right] +
\mathrm{Erf}\left[\frac{\sqrt{2} (p_c+\B \sin\frac{\theta}{2})}{\sqrt{1+2T\nu(\nu+\mu)}}\right]+\right.\label{fid}\\
&2\mathrm{Re}\left\{\frac{\mathrm{Erf}\left[p_c\sqrt{\frac{2(\mu-\nu(T+e^{i\theta}
R))}{\mu+\nu(T-e^{i\theta}
R)}}\right]}{\sqrt{\mu^2-\nu^2(T+e^{{i\theta}}R)^2}}\right.
\left.\left.\exp\left[\frac{2i\A\sin\frac{\theta}{2}\left(
\A e^{\frac{i\theta}{2}}R-2\B\sqrt{T}\right) }{\mu+\nu(T+e^{i\theta} R)}
-\frac{2i\B^2 \sin\frac{\theta}{2}\left(\mu^2 e^{-\frac{i\theta}{2}}-
\nu^2\left(e^{\frac{i\theta}{2}}T+e^{\frac{3i\theta}{2}} R\right)+
i\mu\nu T\sin\frac{\theta}{2} \right)}{\mu^2-\nu^2(T+e^{{i\theta}}R)^2}\right]
\right\}\right\}.\n
\ea
Squeezing enters the above formula through $\mu=\cosh r$ and
$\nu=-\sinh r$, characterizing, together with $\A$,
the initial qumode state. The parameter $\theta$ determines the atom-light
interaction, loss is introduced through $T$ and $R=1-T$, i.e.,
the beamsplitter transmission/reflection coefficients, and $\B$ describes
the (re-)amplification via the displacement operator. The corresponding
probability of success is given by:
\ba
P_s= 
\frac{1}{4}\big[\mathrm{Erf}(c_{00})+\mathrm{Erf}(c_{01})
 +\mathrm{Erf}(c_{10})+\mathrm{Erf}(c_{11})\big],
&\mathrm{ where}&  c_{nm}=\frac{\sqrt{2} \left\{p_c-
(-1)^{n}\left[\B
\sin\frac{\theta}{2}+\sqrt{T} \A \sin(m
\theta)\right]\right\}}{\sqrt{1+ 2T \nu(\nu+\mu \cos(2m\theta))
\label{Ps}
}}\,,
\ea 
\end{widetext}
and $n,m=0,1$.

}

\end{document}